\documentclass[aps,prl,twocolumn,floatfix,showpacs]{revtex4}
\usepackage{epsfig}
\usepackage{graphicx}
\usepackage{color}
\def\be{\begin{equation}}
\def\ee{\end{equation}}
\newcommand{\la}{\langle}
\newcommand{\ra}{\rangle}
\usepackage{subfigure}

\begin{document}

\title{\bf Strong impact of light induced conical intersections\\ on the spectrum of diatomic molecules}

\author{Milan \v{S}indelka$^1$, Nimrod Moiseyev$^1$, and Lorenz S.~Cederbaum$^2$}

\affiliation{$^1$ Schulich Faculty of Chemistry,
 Department of Physics, and Minerva Center of Nonlinear Physics in Complex Systems,
             Technion - Israel Institute of Technology, Haifa 32000, Israel\\
             $^2$ Theoretische Chemie, Physikalisch-Chemisches Institut, Universit\"{a}t Heidelberg, D-69120, Heidelberg, Germany}

\begin{abstract}
We show that dressing of diatomic molecules by running laser waves gives rise to conical intersections (CIs).
Due to presence of such CIs, the rovibronic molecular motions are strongly coupled. A pronounced impact of the CI on the spectrum of
$Na_2$ molecule is demonstrated via numerical calculation for weak and moderate laser intensity, and an experiment is suggested on this basis.
The position of the light
induced CI and the strength of its non-adiabatic couplings can be chosen by changing the frequency and intensity of the used running laser
wave. This offers new possibilities to control the photo-induced rovibronic molecular dynamics.
\end{abstract}


\maketitle

The spectrum of diatomics in weak laser fields is usually interpreted in terms of the Franck-Condon (FC) principle (see e.g.~Ref.~\cite{Atkins}).
If the laser light becomes more intense, the common approach is to take into consideration the two relevant electronic potential energy curves
(PECs) coupled by the laser, and to account for rotational transitions via the selection rule $\Delta J = \pm 1$
(see e.g.~Refs.\cite{Roland,Adi-Natan,Atabek} and citations therein). For even stronger laser intensities, molecular rotations must be
accounted for, as done e.g.~in Ref.~\cite{Charron}.

For standing laser waves we have shown \cite{ci-paper-1} that laser light induces conical intersections (CIs) which couple the center of
mass motion with the internal rovibronic degrees of freedom. As we shall discuss in the present work, laser induced CIs are formed even for running laser waves. It is
well known \cite{ci-book-1} that for the formation of a CI one needs at least two nuclear degrees of freedom whose changes affect the electronic
wavefunction. These are not available for a free diatomic molecule. However, as we shall demonstrate below, the rotation becomes the missing
degree of freedom to allow for the formation of CIs for diatomics in the presence of a laser field. Since $Na_2$ is a favorite and well studied
molecule of laser physics \cite{Na-2-papers-Gerber,Na-2-resolution}, we choose it as our example. The situation is best illustrated in Fig.1.
The top panel shows the potential energy curves of the $X\,^1\Sigma_g^+$ and $A\,^1\Sigma_u^+$ electronic states of $Na_2$. The vertical arrow
labelled by $\hbar\omega_L$ indicates the transition between these two electronic states by a weak laser field in accordance with the FC theory.
Fig.1b depicts the adiabatic potentials obtained by dressing the just mentioned two PECs by the laser following the standard approach of
Refs.\cite{Roland,Adi-Natan,Atabek} referred to above. In Fig.1c we plot the laser induced CI resulting from our approach as discussed below.

\begin{figure}[ht]
%\centering
\subfigure{
\includegraphics[scale=0.50,angle=270]{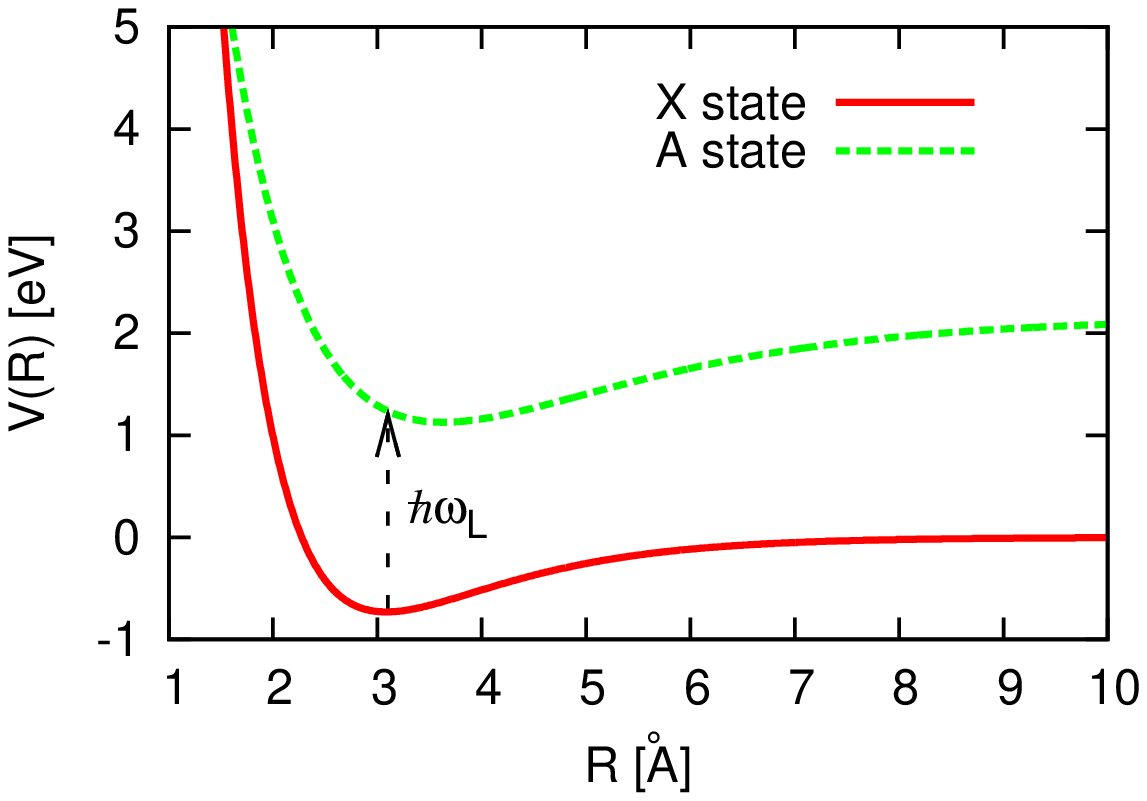}
%\label{fig:subfig1}
}\\
\subfigure{
\includegraphics[scale=0.50,angle=270]{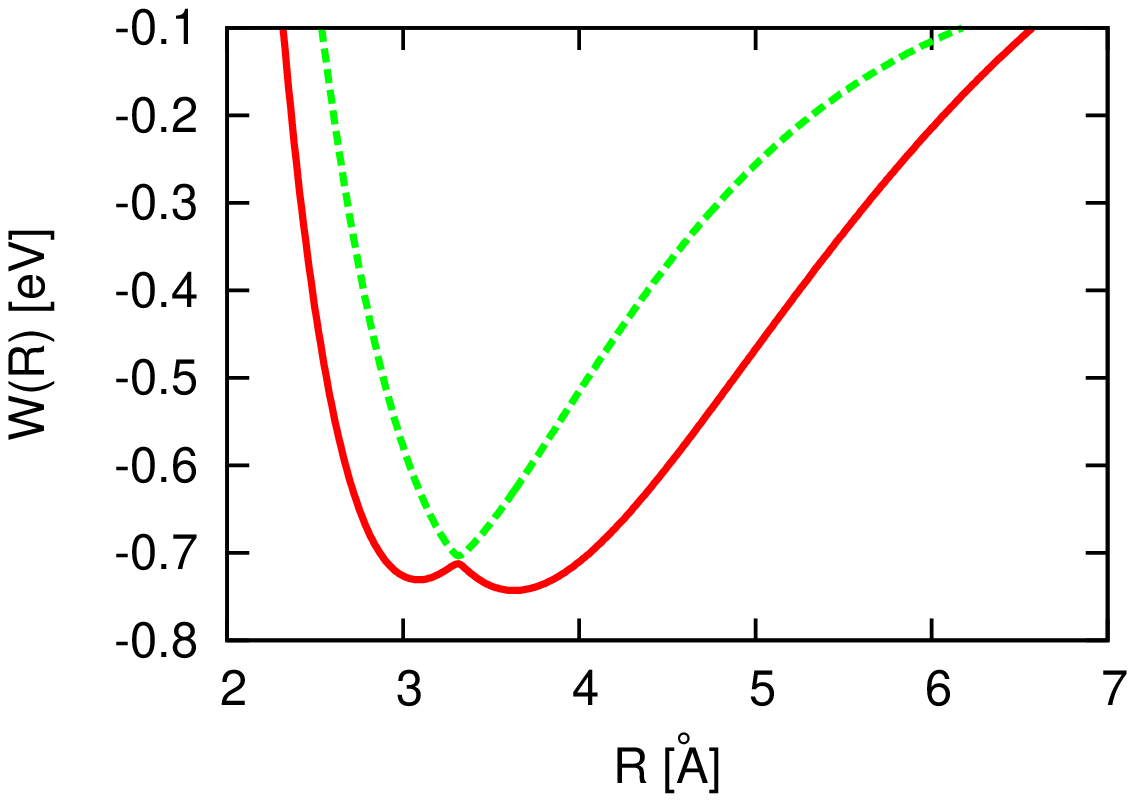}
%\label{fig:subfig2}
}\\
\subfigure{
\includegraphics[scale=0.55,angle=270]{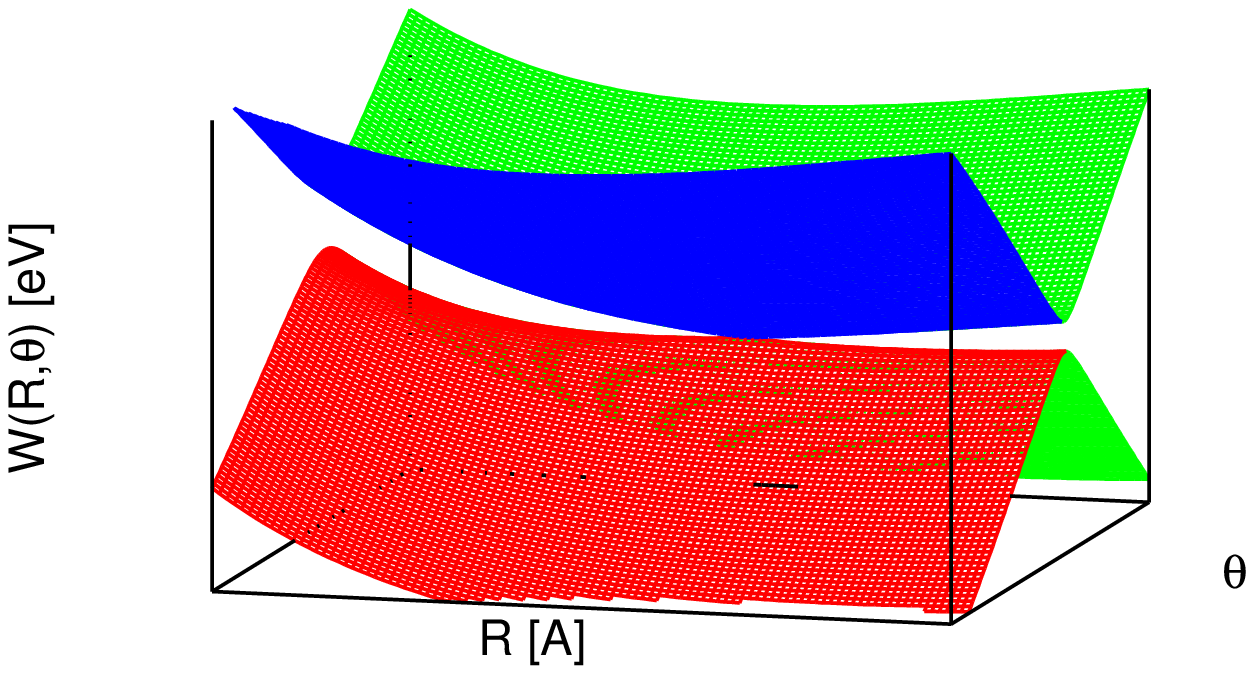}
%\label{fig:subfig3}
}
\label{Fig1}
 \vspace*{-0.40cm}
\caption[bla]{
Electronic potential energies of the $Na_2$ molecule
dressed by laser light, pertaining to different levels of theory
discussed in the text. Top: The field free PECs. Middle: The two lowest lying dressed adiabatic potentials for a
field intensity of $3\cdot10^8$ W/cm$^2$. Bottom: Conical intersection induced by a running laser wave.
For visualization reasons we have used here a rather strong
laser intensity of $10^{14}$ W/cm$^2$. Note however that the conical intersection exists
for any nonzero field strength. \vspace*{-0.60cm}}
\end{figure}

The presence of CIs is well known in polyatomics to exert an enormous impact on the molecular dynamics
\cite{ci-book-1,Lenz-Advances}. We will show here for the first time that this is also the case for the CI induced by running laser waves in diatomics.

Let us briefly describe now the underlying theoretical background.
Using the same formalism and assumptions as in \cite{ci-paper-1} one can derive the following Hamiltonian for rovibronic molecular motions:
\begin{eqnarray} \label{H-TOT}
   {\bf H} & = & \left( - \frac{\hbar^2}{2\mu}\frac{\partial^2}{\partial R^2} \, + \, \frac{{\bf L}_{\theta\varphi}^2}{2 \mu R^2} \right) \otimes \left( \matrix{ 1 & 0 \cr 0 & 1 } \right) \\
   & + & \left( \matrix{ V_X(R) & ({\cal E}_0/2) \, d(R) \, \cos\theta \cr ({\cal E}_0/2) \, d(R) \, \cos\theta & V_A(R) - \hbar \omega_L } \right)
   \; \; . \nonumber
\end{eqnarray}
Here, $R$ and $(\theta,\varphi)$ are the molecular vibrational and rotational coordinates, respectively, $\mu$ stands for the reduced mass of
the diatomic, and ${\bf L}_{\theta\varphi}$ represents the angular momentum operator of the nuclei. The center of mass motion has been
eliminated via the usual adiabatic separation approach which is physically adequate for the case of running waves.
Subscripts $X$ and $A$ refer to the
two electronic states coupled by the laser (whose frequency is $\omega_L$ and amplitude ${\cal E}_0$). To be specific we assume that these
two electronic states possess $\Sigma$ type symmetry, such that the molecule-light coupling term is proportional to
$d(R) \, \cos\theta$, where $d(R)$ is the corresponding dipole transition matrix element in the body fixed frame.
As discussed in detail in our previous work \cite{ci-paper-1}, the potential matrix of the Hamiltonian (\ref{H-TOT}) gives rise to a conical
intersection whenever the conditions $V_X(R) = V_A(R) - \hbar \omega_L$ and $\theta=\pi/2$ are met. This situation is shown explicitly in Fig.1c.

The conventional Hamiltonian $\tilde{\bf H}$ widely used for diatomics in laser fields (see e.g.~Refs.\cite{Roland,Adi-Natan,Atabek}) is obtained from (\ref{H-TOT}) by
replacing $\cos\theta$ by its optical transition matrix element
\be \label{tilde-H-def}
   \hspace*{-0.40cm} \la 0 | \cos\theta | 1 \ra = \int_{0}^{\pi} P_0(\cos\theta) \cos\theta P_1(\cos\theta) \sin\theta \, d\theta \, .
\ee
Here, $| l \ra \equiv P_l(\cos\theta)$ are the Legendre polynomials.

To illustrate the impact of the light induced CI on observable quantities we discuss the $Na_2$ molecule,
considered above in Fig.1. The PECs and the transition dipole functions
pertaining to the $X\,^1\Sigma_g^+$ and $A\,^1\Sigma_u^+$ electronic states are taken from \cite{Magnier} and
the laser wavelength is $\lambda=663$ nm.

Solution of the eigenvalue problem of the Hamiltonian (\ref{H-TOT}) provides the
quasi-energy eigenvalue spectrum. These quasi-energy eigenvalues and eigenfunctions are calculated numerically by representing the
${\bf H}$ as a matrix in the basis set of the free molecular rovibronic states. An energy truncation criterion is used to reduce the number of basis 
functions to a manageable size. Numerical convergence with respect to this energy cutoff has been checked and achieved.
We shall thus hereafter refer to the obtained numerical results as "exact".

In the case of weak laser intensities one expects that the second order perturbation theory (PT2) in ${\cal E}_0$ will be adequate for the energy
levels. Indeed, if we use the common approach (which we call in the following the "no CI" case), we find this expectation to be fulfilled.
The situation changes dramatically when full account of the CI is taken. This is apparent in Fig.2, where the deviations of the PT2 results from
the exact ones are shown. Breakdown of PT2 in the vicinity and above the energetic location of the CI is clearly demonstrated.
\begin{figure}[ht]
\includegraphics[scale=0.50,angle=270]{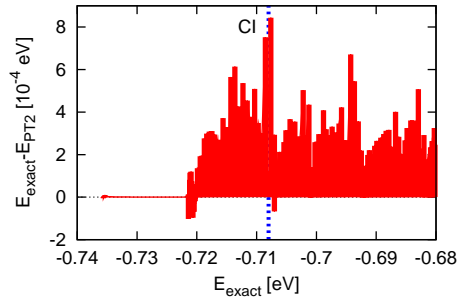}
\vspace*{-0.40cm}
\caption{Breakdown of the second order perturbation theory in the vicinity of the energy of the CI and above it.
         The position of the CI is marked by a vertical dotted line. \vspace*{-0.60cm} }
\end{figure}

In Fig.2 all the lowest energy levels of the Hamiltonian (\ref{H-TOT}) have been considered. To make contact with a possible experiment, we study
below the absorption spectrum of $Na_2$ where the levels are weighted by their intensities. We discuss three types of spectra: The FC spectrum obtained by
employing the standard FC principle, the "no CI" spectrum determined by the standard approach, and the "exact" spectrum following from (\ref{H-TOT}).
The Golden rule expression for the "exact" spectral intensity is given by $|\wp(E)|^2$, where
\begin{eqnarray} \label{spectral-intensity}
   \wp(E) & = & \la \Psi_{0}^X(R,\theta) \, | \, d(R) \, \cos\theta \, | \, \Psi_E^A(R,\theta) \ra \nonumber\\
          & + & \la \Psi_{0}^A(R,\theta) \, | \, d(R) \, \cos\theta \, | \, \Psi_E^X(R,\theta) \ra \; .
\end{eqnarray}
%\be \label{spectral-intensity}
%   \Bigl| \, \la \phi_{(\nu=0,J=0)}^X(R) \, | \, d(R) \, \cos\theta \, | \, \Psi_E^A(R,\theta) \ra \, \Bigr|^2 \;\; .
%\ee
%{\it Shall we say where it comes from ?}\\
Here, $\vec{\Psi}_{0}(R,\theta)=(\Psi_{0}^X(R,\theta),\Psi_{0}^A(R,\theta))^T$ stands for that particular eigenstate of the Hamiltonian (\ref{H-TOT}) which
reduces to the ground state of the free diatomic molecule in the limit of ${\cal E}_0 \to 0$. Similarly,
$\vec{\Psi}_{E}(R,\theta)=(\Psi_{E}^X(R,\theta),\Psi_{E}^A(R,\theta))^T$ represents any other eigenstate of the Hamiltonian (\ref{H-TOT}),
with quasi-energy $E$. The physical picture behind the just defined spectral intensity is as follows. First, a sample of
free diatomic molecules (cold enough to be predominantly in their ground states) is dressed by an adiabatically switched on CW laser pulse with
central frequency $\omega_L$. This leads via an adiabatic passage to the dressed molecular state $\vec{\Psi}_{0}(R,\theta)$ described above.
Subsequently, the system is probed by another laser which facilitates dipole transitions from $\vec{\Psi}_{0}(R,\theta)$ to $\vec{\Psi}_{E}(R,\theta)$
according to the standard Golden rule theory as described by Eq.~(\ref{spectral-intensity}).
The FC spectrum is obtained from (\ref{spectral-intensity}) by identifying the initial state $\vec{\Psi}_{0}(R,\theta)$ with the ground state of
a free diatomic molecule, and by choosing the final states $\vec{\Psi}_{E}(R,\theta)$ to be the eigenstates of the $A\,^1\Sigma_u^+$ state
in absence of the laser.
It is known \cite{Atkins} that the FC spectrum becomes precise in the weak field limit of ${\cal E}_0 \to 0$. For the "no CI" case, the spectrum is
obtained from (\ref{spectral-intensity}) by substituting for $\vec{\Psi}_{0}(R,\theta)$ and $\vec{\Psi}_{E}(R,\theta)$ the corresponding eigenstates of the
modified approximative Hamiltonian $\tilde{\bf H}$ which was defined above using Eq.~(\ref{tilde-H-def}).

\begin{figure}[ht]
%\centering
\subfigure{
\includegraphics[scale=0.38,angle=270]{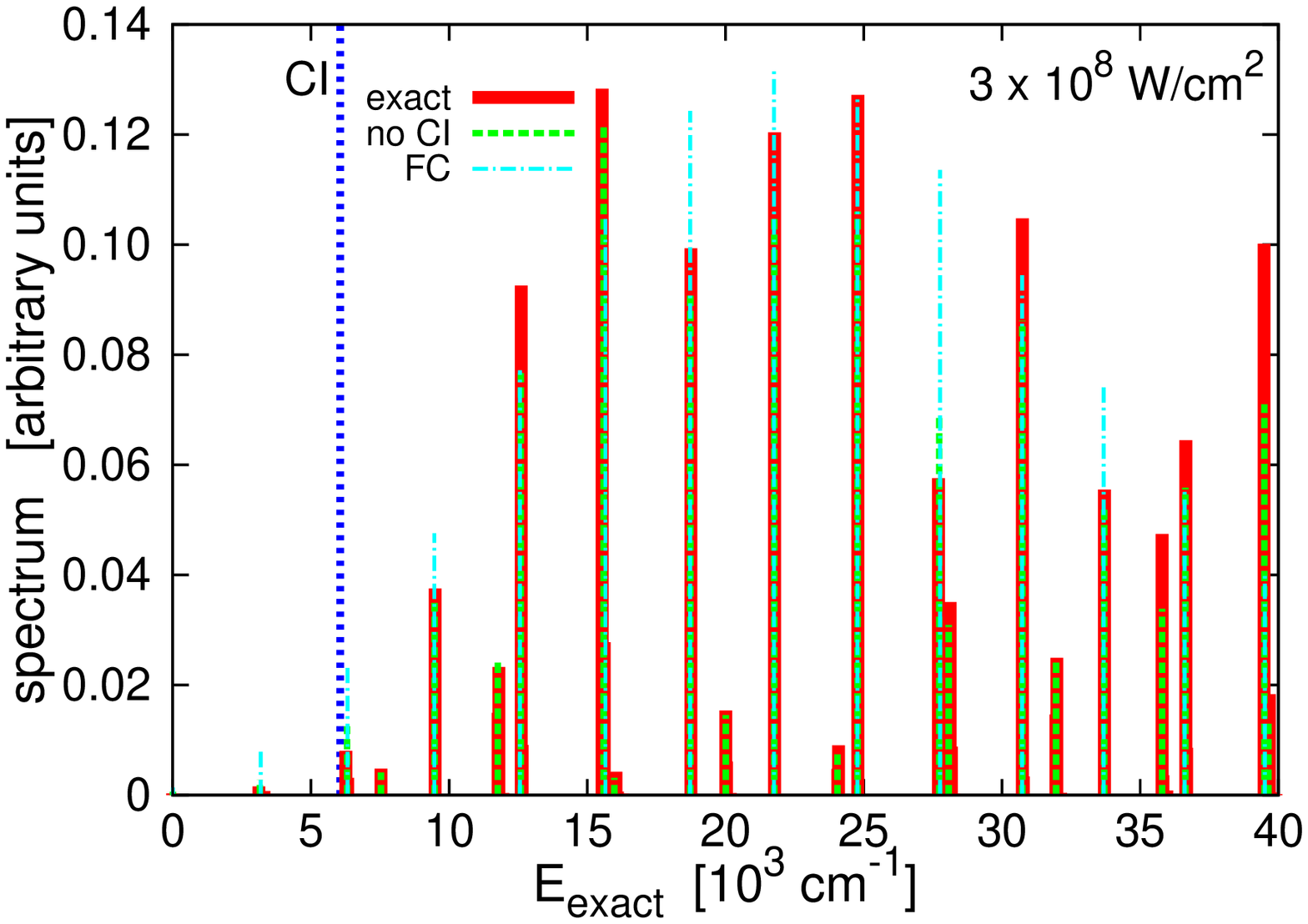}
%\label{fig:subfig1}
}\\
%\subfigure[Absorption spectrum of $Na_2$ dressed by a running laser wave ($\lambda=663$ nm, intensity $10^{11}$ W/cm$^2$.)]{
%\includegraphics[scale=0.35,angle=270]{Fig3b-new.ps}
%%\label{fig:subfig2}
%}\\
\subfigure{
\includegraphics[scale=0.38,angle=270]{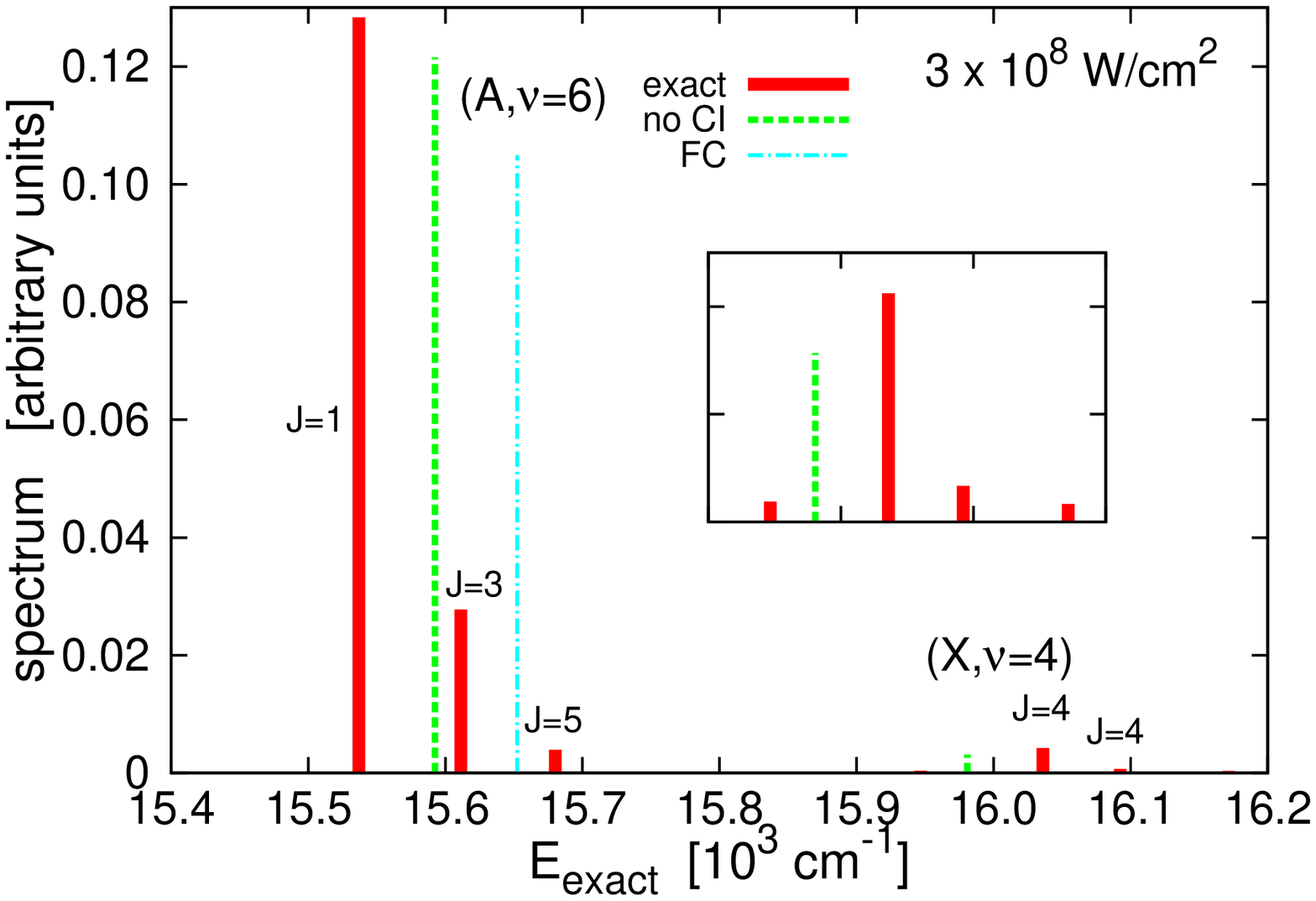}
%\label{fig:subfig3}
}
\label{Fig3}
\vspace*{-0.40cm}
\caption[bla]{
Rovibronic absorption spectra of $Na_2$ dressed by a running laser wave ($\lambda=663$ nm, intensity $3 \cdot 10^8$ W/cm$^2$).
The calculations are performed at different levels of theory (FC, "no CI", exact) discussed in the text. Top: overview of the spectrum.
Bottom: Zoom into the range of $15400$ and $16200$ cm$^{-1}$ (with an inset showing the range between $15900$ and $16200$ cm$^{-1}$
in an extended scale). Fingerprints of CI are clearly demonstrated by pronounced deviation of the
exact spectrum from the FC and "no CI" approximations. \vspace*{-0.50cm} }
\end{figure}

The computed spectra are depicted in Fig.3 for a weak laser intensity of $3 \cdot 10^8$ W/cm$^2$. Fig.3a gives an overview of the calculated
transition intensities. For most lines the FC approximation describes the spectrum qualitatively well. Nevertheless, clear deviations from the
"no CI" spectrum are seen for most lines, with the intensity of the FC lines being higher (in particular for lines with energies about $18000$
and $28000$ cm$^{-1}$ above the ground state). Moreover, many FC lines are seen to split into two in the "no CI" spectrum.
Interestingly, even at this weak laser field one can see obvious differences between the "no CI" and the "exact" spectra. The richness
of the "exact" spectrum becomes apparent on an expanded energy scale. In Fig.3b we show the vicinity of the FC line between
$15400$ and $16200$ cm$^{-1}$. While there is only a
single FC line (characterized by the electronic state $A$, and $\nu=6$, $J=1$), we observe two lines in the "no CI" spectrum. The
second weaker line is mainly the vibrational level ($\nu=4$, $J=0$) of the lower electronic state $X$ and has acquired its intensity by borrowing it
from the more intense line ($\nu=6$, $J=1$) of the $A$ electronic state. This borrowing of intensity is a well known non-adiabatic effect
\cite{Lenz-Advances}.
Interestingly, the CI gives rise to further substantial splittings. Each of the "no CI" peaks splits into essentially three lines. For the weak
laser intensity at hand, the splitting is mostly due to rotations. The assignment of the exact levels in Fig.3b refers to the leading configuration
in the expansion of the corresponding eigenstates in terms of the free molecular states. In reality, several (or even many) configurations contribute.
For example, the state associated with the most intense borrowing line in the spectrum (labelled by $X$, $\nu=4$, $J=4$) has a weight on the excited
$A$ electronic state of $20\%$. The averaged rotational quantum number $J_X$ on the $X$ electronic state equals to $J_X \pm \Delta J_X = 1.8 \pm 1.8$
where $\Delta J_X$ is the standard deviation. Analogously, on the $A$ electronic state, $J_A \pm \Delta J_A = 0.4 \pm 1.2$.
%{\it Shall we choose another section of the spectrum and show it in detail as in Fig.3b ? Namely that part where the borrowing peaks are
%more pronounced in magnitude. }

We note that the absorption profiles depicted in Fig.3 require spectroscopic resolution better than $50$ cm$^{-1}$ which can be obtained
in the lab (see, e.g., Ref.~\cite{Na-2-resolution}). In supersonic beam experiments \cite{Na-2-papers-Gerber} the predicted phenomenon should be
possible to observe. Fig.3 may thus serve as a proposal of a future experiment aimed at demonstrating the impact of light induced CIs
on molecular rovibronic spectra even for rather weak fields.

Let us discuss now the situation encountered at the larger field intensity of $10^{11}$ W/cm$^2$. Here, all the molecular degrees of freedom are
strongly coupled, and the corresponding eigenstates of the Hamiltonian (\ref{H-TOT}) cannot be assigned to their field free
counterparts. This complicates the use of the Golden rule formula (\ref{spectral-intensity}), as we are unable to populate the initial state
$\vec{\Psi}_0(R,\theta)$ via an adiabatic passage. In order to gain an insight into the nature of the obtained dressed molecular states, we
instead consider the projection
$
   P(E) = |\la \Psi_E^X(R,\theta) | \phi_{\nu=0,J=0}^X(R) \ra|^2
$
of each dressed state $\vec{\Psi}_E(R,\theta)$ on the field free molecular ground state $\phi_{\nu=0,J=0}^X(R)$. The quantity $P(E)$ possesses a
clear physical meaning. It is the probability of populating the dressed state $\vec{\Psi}_E(R,\theta)$ when the laser pulse is imposed
abruptly on the sample of free diatomic molecules residing in their ground states $\phi_{\nu=0,J=0}^X(R)$.
For the considered field intensity, pulse shaping techniques allow the
production of nearly any square laser pulse, depending on the availability of
a source that can cover the required bandwidth \cite{pulse-shaping}.
\begin{figure}[ht]
\includegraphics[scale=0.35,angle=270]{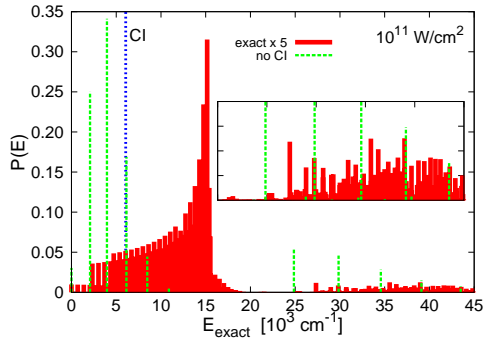}
\vspace*{-0.40cm}
\caption{ Probability $P(E)$ of populating different dressed molecular states when an intense laser light
          ($\lambda=663$ nm, intensity $10^{11}$ W/cm$^2$) arrives abruptly to a sample of free $Na_2$ molecules in their ground states.
          For visualization reasons, the exact spectrum is multiplied by factor 5. An inset shows a zoom into the region between
          20000 and 40000 cm$^{-1}$, where a completely irregular pattern of the profile of $P(E)$ is evident.
          \vspace*{-0.60cm} }
\end{figure}

Fig.4 shows the obtained profile of $P(E)$. % for the laser intensity of $10^{11}$ W/cm$^2$.
% The situation changes dramatically when increasing the laser intensity. In Fig.3b we show all three spectra for the laser intensity of
% $10^{11}$ W/cm$^2$. The FC spectrum is, of course, the same as discussed above in Fig.3a. Now, obviously, this spectrum differs substantially
% from the other spectra, and need not be discussed further. In contrast to the weak field case, the "no CI" spectrum exhibits lines which are more
% intense than the FC ones. 
Surprisingly, it turns out that the "no CI" spectrum has no resemblance whatsoever with the exact spectrum. The "no CI" spectrum has become
very sparse and the intensity of its lines is typically more than one order of magnitude larger than that of the lines of the extremely dense
exact spectrum. This sparseness is
a strong indication that at this laser field strength the non-adiabatic effects in the "no CI" model have reduced enormously.
%{\it Analysis: how is the mixing of EL, $\nu,J$.}
The high density of lines in the exact spectrum, on the other hand, is a signature of the singular non-adiabatic
coupling induced by a strong CI \cite{Lenz-Advances}. In particular, the spectrum above the energy of about 20000 cm$^{-1}$
%of the CI (indicated by the blue vertical line in the figure)
exhibits a typical
irregular pattern like the spectra of CIs in polyatomics which have been shown to behave like the spectra of random matrix ensembles
\cite{Lenz-random-matrices}.% {\it Should we show a zoom into this energy region ?}
This irregularity reflects the complete mixing of multiple configurations in the respective eigenstates. Our numerical analysis shows that both the X and A
electronic states are essentially equally populated in {\it all} the eigenstates above the CI energy. Similarly, also the field free vibrational as well as
rotational levels are enormously mixed. For example, for the lines between $40000$ and $45000$ cm$^{-1}$ one typically finds
$J_X \pm \Delta J_X = 20 \pm 20$ and $J_A \pm \Delta J_A = 25 \pm 25$, while
$\nu_X \pm \Delta \nu_X = 3 \pm 3$ and $\nu_X \pm \Delta \nu_X = 5 \pm 5$. In all cases, the magnitudes of the calculated
averages are roughly equal to their standard deviations.

We have shown that even diatomic molecules exhibit CIs. These are induced by running laser waves,
and live in the space of the molecular rotational and vibrational degrees of freedom. We have demonstrated
that these CIs have a strong impact on the molecular spectrum even for weak laser fields. For intermediate laser fields this
impact becomes dramatic and the resulting spectrum has no resemblance with those computed by standard approaches. The coupling of two electronic
states via the light induced conical intersection is substantially and sometimes even dramatically stronger than without it, and consequently,
one can apply rather weak fields to achieve strong coupling between these two states. If needed, e.g., when the field is strong and other
electronic states are close by, one can easily include more electronic states in the formalism. Spectra of field free molecules, like that of
copper trimer \cite{copper-trimer} or sodium trimer \cite{sodium-trimer}
have been shown to bear the signature of the Berry phase in the electronic wave function due to CIs \cite{Mead-Truhlar}.
The same classic consequence of the Berry phase in the electronic wave function is also found here for light induced CIs.
In sharp contrast to field free polyatomic molecules where the CI is given by nature, the energetic position of the
CI discussed here can be controlled by the laser frequency and the strength of its non-adiabatic coupling
by the laser intensity. Finally, we stress that for polyatomics there are, of course, also laser induced CIs. It is anticipated that their
interplay with the CIs given by nature will lead to a wealth of new phenomena.

The authors are thankful for fruitful discussions with A.~Zohar and A.~Fleischer.
ISF 96/07, ERC, and DFG grants are acknowledged for their support.

%M.~\v{S}.~acknowledges financial support by an Advanced Grant from the European Research Council (ERC), N.~M.~by the Israel
%Science Foundation (Grant No.~96/07), and L.~S.~C.~by the Deutsche Forschungsgemeinschaft.

\end{document}